\documentclass[a4paper,11pt]{article}
\pdfoutput=1

\usepackage{jcappub}

\usepackage[T1]{fontenc} 

\usepackage{epsfig}
\usepackage{latexsym}
\usepackage{graphicx}
\usepackage{epstopdf} 
\usepackage[justification=RaggedRight]{caption}
\usepackage{amssymb}
\usepackage{subfig}
\usepackage{color}
\usepackage{multirow}
\usepackage{color}
\usepackage{rotating}
\usepackage{ifthen}
\usepackage{epsfig}
\usepackage{booktabs} 
\usepackage{tikz}
\usetikzlibrary{arrows,shapes}
\usetikzlibrary{trees}
\usetikzlibrary{matrix,arrows} 				
\usetikzlibrary{positioning}				
\usetikzlibrary{calc,through}				
\usetikzlibrary{decorations.pathreplacing}  
\usepackage{pgffor}							

\usetikzlibrary{decorations.pathmorphing}	
\usetikzlibrary{decorations.markings}
\tikzset{
    vector/.style={decorate, decoration={snake}, draw},
	provector/.style={decorate, decoration={snake,amplitude=2.5pt}, draw},
	antivector/.style={decorate, decoration={snake,amplitude=-2.5pt}, draw},
    fermion/.style={draw=black, postaction={decorate},
        decoration={markings,mark=at position .55 with {\arrow[draw=black]{>}}}},
    fermionbar/.style={draw=black, postaction={decorate},
        decoration={markings,mark=at position .55 with {\arrow[draw=black]{<}}}},
    fermionnoarrow/.style={draw=black},
    gluon/.style={decorate, draw=black,
        decoration={coil,amplitude=4pt, segment length=5pt}},
    scalar/.style={dashed,draw=black, postaction={decorate},
        decoration={markings,mark=at position .55 with {\arrow[draw=black]{>}}}},
    scalarbar/.style={dashed,draw=black, postaction={decorate},
        decoration={markings,mark=at position .55 with {\arrow[draw=black]{<}}}},
    scalarnoarrow/.style={dashed,draw=black},
    electron/.style={draw=black, postaction={decorate},
        decoration={markings,mark=at position .55 with {\arrow[draw=black]{>}}}},
	bigvector/.style={decorate, decoration={snake,amplitude=4pt}, draw},
}

\tikzstyle{block} = [draw, rectangle, 
    minimum height=3em, minimum width=6em]

\usepackage{hyperref} 
\usepackage{hyperref}
\hypersetup{
    colorlinks,%
    citecolor=blue,%
    filecolor=blue,%
    linkcolor=blue,%
    urlcolor=blue
}
\def\be{\begin{equation}}
\def\ee{\end{equation}}

\newcommand{\bwt}{\begin{widetext}}
\newcommand{\ewt}{\end{widetext}}
\newcommand{\bdm}{\begin{displaymath}}
\newcommand{\edm}{\end{displaymath}}
\newcommand{\bea}{\begin{eqnarray}}
\newcommand{\eea}{\end{eqnarray}}

\newcommand{\beqra}{\begin{eqnarray}}
\newcommand{\eeqra}{\end{eqnarray}}
\newcommand{\beq}{\begin{equation}}
\newcommand{\eeq}{\end{equation}}

\subheader{LPT-Orsay-14-90}

\title{3.55 keV line in Minimal Decaying Dark Matter scenarios}

\author[a]{Giorgio Arcadi,}

\author[b]{Laura Covi}

\author[b]{and Federico Dradi}

\affiliation[a]{Laboratoire de Physique Th\'eorique 
Universit\'e Paris-Sud, F-91405 Orsay, France}
\affiliation[b]{Institute for Theoretical Physics, 
Georg-August University G\"ottingen, 
Friedrich-Hund-Platz~1, G\"ottingen, D-37077 Germany}

\abstract{
We investigate the possibility of reproducing the recently reported $3.55\,\mbox{keV}$ line in some simple decaying dark matter scenarios. 
In all cases a keV scale decaying DM is coupled with a scalar field charged under SM gauge interactions and thus capable of pair production 
at the LHC. We will investigate how the demand of a DM lifetime compatible with the observed signal, combined with the requirement of 
the correct DM relic density through the freeze-in mechanism, impacts the prospects of observation at the LHC of the decays of the scalar field. 
}

\begin{document} 
\maketitle
\flushbottom

\section{Introduction}

The identification of the Dark Matter component of the Universe is one of the most important puzzles in modern astrophysics and particle physics. 
Although conventional paradigms rely on stable particle states, there are no a-priori arguments against decaying dark matter candidates, 
provided that their lifetimes largely exceed the age of the Universe. In order to satisfy this requirement, the couplings of the DM with ordinary matter 
are too suppressed to allow Direct Detection of DM in experiments like e.g. XENON~\cite{Aprile:2012nq,Aprile:2013doa}, LUX~\cite{Akerib:2013tjd}. 
On the contrary there are rather promising prospects on Indirect Detection, in cosmic rays, of the DM decays occurring at present times. 
Already strong limits on the DM lifetime have been set on a broad variety of decay products by observations like those performed by 
Fermi~\cite{Zaharijas:2012dr,Ackermann:2011wa,Morselli:2012xra}, AMS~\cite{Aguilar:2013qda} and XMM/CHANDRA~\cite{Herder:2009im}.

$\gamma/X$-rays are among the most promising signatures of DM Indirect Detection. Indeed DM decays (and annihilations) can give rise to sharp 
peaks (``lines'') which can be hardly accounted for by astrophysical sources. A spectral feature of this kind has been recently identified in the combined
spectrum of a large set of X-ray galaxy clusters~\cite{Bulbul:2014sua} as well as in the combined observation of the Perseus Cluster and the M31 
galaxy~\cite{Boyarsky:2014jta}. This signal can be accounted for by a rather broad variety of models of decaying DM~\cite{Abazajian:2014gza,Baek:2014qwa,Tsuyuki:2014aia,Okada:2014zea,Cline:2014eaa,Modak:2014vva,Allahverdi:2014dqa,Robinson:2014bma,Rodejohann:2014eka,Haba:2014taa,Finkbeiner:2014sja,Jaeckel:2014qea,Lee:2014koa,Cicoli:2014bfa,Kong:2014gea,Choi:2014tva,Conlon:2014wna,Bomark:2014yja,Demidov:2014hka,Nakayama:2014ova,Chiang:2014xra,Shuve:2014doa,Kolda:2014ppa,Dutta:2014saa,Queiroz:2014yna,Geng:2014zqa,Cline:2014kaa,Krall:2014dba,Falkowski:2014sma} as well as annihilating DM~\cite{Baek:2014poa,Dudas:2014ixa} 
although an astrophysical/instrumental origin is still feasible~\cite{Jeltema:2014qfa,Carlson:2014lla} as the line has not been seen in stacked
spheroidal galaxies or galaxy groups \cite{Malyshev:2014xqa, Anderson:2014tza}.

A very simple and rather predictive decaying Dark Matter scenario has been proposed and studied in~\cite{Arcadi:2013aba,Arcadi:2014tsa}. 
Here a SM singlet Majorana fermion is the Dark Matter candidate and it is coupled with a scalar field charged under (at least some of) the gauge interactions 
of the Standard Model and Standard Model fields. 
In its simplest realization this model can be described by only four parameters: the masses of the DM and of the scalar field and two Yukawa-type couplings 
$\lambda$ and $\lambda^{'}$ for the scalar field. 
The correct DM relic density is ensured by a combination of the freeze-in and SuperWIMP mechanisms which are effective at the low values of the 
couplings $\lambda$ and $\lambda^{'}$ compatible with constraints from Indirect Detection. An hypothetical detection of the DM decays can in such
a simple setting be related  to LHC signals associated to the LHC production of the charged scalar field with the latter being long-lived or even detector-stable, 
thus producing peculiar signatures in the form of displaced decays and/or disappearing tracks.

This kind of relation has been established  in~\cite{Arcadi:2013aba,Arcadi:2014tsa} for rather massive, at least at the GeV scale, DM candidates, for 
which three body tree-level decay processes of the DM into SM fermions are kinematically allowed. 
In this work we want to investigate whether a similar interplay between collider and DM indirect detection can be established for DM masses at the keV scale, 
for which only two-body, one-loop induced, decays into a neutrino and a photon are possible, thus reproducing in the model the X-ray line signal.
We will as well discuss two extensions/modifications of the scenario, allowing for further couplings and fields.
In the first case we will consider the case in which the DM decays into a photon and a new SM singlet, rather than the neutrino. 
This new state is required to be extremely light, $~O(\mbox{eV})$, and can then contribute to the number of light species $N_{\rm eff}$ which are 
probed by CMB experiments. We will finally consider the case in which the DM plays the role of a sterile neutrino. 
Although in this case the line signal is accounted for by the mixing with SM neutrinos, the presence of the scalar field is still relevant since 
it provides an additional DM generation mechanism and lightens some of the cosmological tensions characteristic of this kind of scenarios.

The paper is organized as follows. In the next section we will briefly review the minimal decaying DM model and investigate the impact on the parameter 
space of the combined requirement of the agreement with the X-ray signal and of the correct DM relic density. In section 3 and 4, we will then discuss the 
two simple extensions and finally state our conclusions.

\section{Minimal scenario}

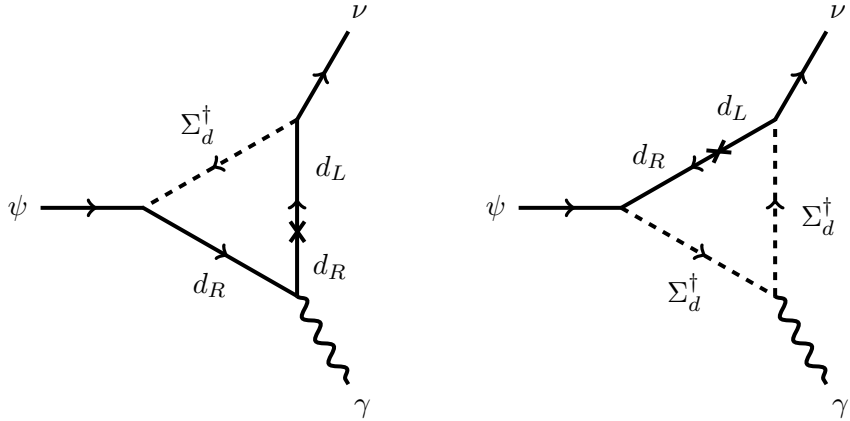
\begin{figure}[!ht]
  \centering
\begin{tikzpicture}[line width=1.5pt,rotate=90,scale=1.5]
	\draw[scalar] (-30:0.9) -- (90:0.9);
	\draw[fermion] (90:0.9) -- (210:0.9);
	\draw[fermion] (210:0.9) -- (-30:0.9);
	\begin{scope}[shift={(-115:.5)},rotate=90]
	\draw (125:.11) -- (-55:.11);
	\draw (55:.11) -- (-125:.11);
	\end{scope}
	\draw[fermion] (-30:0.9) -- (-30:1.8);
	\draw[fermionbar] (90:0.9) -- (90:1.8); 
	\draw[vector] (210:0.9) -- (210:1.8);
	\node at (30:.85) {$\Sigma_d^\dag$};
	\node at (210:2.05) {$\gamma$};
	\node at (-30:2.00) {$\nu$};
	\node at (295:0.83) {$d_L$};
	\node at (235:0.9) {$d_R$};
	\node at (155:0.75) {$d_R$};
	\node at (90:2.00) {$\psi$};
\end{tikzpicture}
\hspace{1cm}
\begin{tikzpicture}[line width=1.5pt,rotate=90,scale=1.5]
	\draw[fermion] (-30:0.9) -- (90:0.9);
	\draw[scalar] (90:0.9) -- (210:0.9);
	\draw[scalar] (210:0.9) -- (-30:0.9);
	\begin{scope}[shift={(5:.5)},rotate=30]
	\draw (125:.11) -- (-55:.11);
	\draw (55:.11) -- (-125:.11);
	\end{scope}
	\draw[fermion] (-30:0.9) -- (-30:1.8);
	\draw[fermionbar] (90:0.9) -- (90:1.8); 
	\draw[vector] (210:0.9) -- (210:1.8);
	\node at (265:.83) {$\Sigma_d^\dag$};
	\node at (210:2.05) {$\gamma$};
	\node at (-30:2.00) {$\nu$};
	\node at (-5:0.89) {$d_L$};
	\node at (55:0.8) {$d_R$};
	\node at (155:0.82) {$\Sigma_d^\dag$};
	\node at (90:2.00) {$\psi$};
\end{tikzpicture}
\caption{DM 2-body decay into $\gamma$ and $\nu$ with the loop induced 
by $\Sigma_d$.}
\label{loopDMdecay}
\end{figure}

A very simple and economic scenario of decaying dark matter has been discussed in~\cite{Arcadi:2013aba} 
(see also~\cite{Garny:2010eg,Garny:2011ii,Garny:2012vt}). A Maiorana DM particle $\psi$, singlet with respect to the SM gauge group, 
features a renormalizable Yukawa type interaction with a scalar field $\Sigma_f$, with not trivial SM charges, and a SM fermion, being a lepton 
or a quark according to the assignment of the quantum numbers of $\Sigma_f$:
\begin{equation}
L_{\rm eff} =\lambda \bar{\psi} f\; \Sigma_f^{\dagger} 
+h.c.\,.
\label{eq:DMint}
\end{equation}
In absence of symmetries protecting the DM stability, interactions of the same type are also allowed between $\Sigma_f$ and two SM fermions 
(see~\cite{Arcadi:2014tsa} for a complete list of all the allowed operators), such that the DM can decay into three SM fermions. These kind of processes 
are already severely constrained by antiproton searches~\cite{Garny:2012vt,Fornengo:2013xda} and, more recently, by the measurements by AMS 
of the positron flux and positron fraction~\cite{Ibarra:2013zia}. A photon line can be produced by a DM decay process into a photon and a neutrino, 
induced at one loop by diagrams like the one reported in fig.\,\ref{loopDMdecay}. This kind of process has a sizable branching ratio only when tree-level 
decays into SM fermions are kinematically forbidden. The energy of the photon emitted per DM decay is simply given, by kinematics, by 
$E_\gamma =\frac{m_\psi}{2}$. This fixes the DM mass to about $7\,\mbox{keV}$ if one wants to explain the recently observed X-ray line.

This kind of decay occurs only if the following couplings between the scalar field and SM fermions are present:
\begin{align}
\label{eq:lagrangian}
& L_{\rm eff}=\lambda^{'}  \bar{d}_R \ell_L \Sigma_q+h.c. & \Sigma_q=(3,2,1/3) \nonumber\\
& L_{\rm eff}=\lambda^{'} \bar{\ell}^c_{R}q_L \Sigma_d^{\dagger}+h.c. &  \Sigma_d=(3,1,-2/3) \nonumber\\
& L_{\rm eff}= \lambda^{'} \bar{\ell}^c_R \ell_L \Sigma_e^{\dagger}+h.c. & \Sigma_\ell=(1,2,-1) \nonumber\\
& L_{\rm eff}={\lambda}^{'} \bar{e}_R \ell_L \Sigma_\ell+h.c. & \Sigma_e=(1,1,-2) \, .
\end{align}
Besides the ones shown in (\ref{eq:lagrangian}) other operators coupling the scalar field with two SM fermions can be in general present but have no impact in the 
analysis performed in this work. We remark however that all these operators violate baryon or lepton number and 
very severe constraints from proton decay may arise if both these two quantum numbers are violated. We 
will thus implicitly assume throughout all this work that there is no contemporary presence of operators 
which violate $B$ and $L$ numbers. 

The decay rate of the DM into a neutrino and a photon is given by~\cite{Garny:2011ii}:
\begin{align}
\label{partialDecayWidth}
& \Gamma(\psi\to \gamma \nu)= \frac{e^2 m_\psi^3}{2048 \pi^5} 
\left(\sum_i \frac{m_i}{m_{\Sigma_f}^2}  
\lambda^{'}_i \lambda_i f_1\left(\frac{m_i^2}{m_{\Sigma_f}^2}\right) \right)^2\,,
\nonumber\\
& \mbox{where} \;\; f_1(x)=\frac{1}{1-x} \left[1+\frac{1}{1-x} \ln x \right]
\end{align}
and the sum runs over the fermions flowing into the loop. We notice that the decay rate depends on the mass of the SM fermion in the loop since a chirality flip 
in the internal fermion line is required. Unless particular hierarchies in the couplings $\lambda$ and $\lambda^{'}$ with respect to the fermion flavors are assumed 
the DM decay rate is thus mostly sensitive to the couplings of $\Sigma_f$ with third generation fermions. From now on, unless differently stated, we will assume the 
couplings $\lambda$ and $\lambda^{'}$ flavor universal and keep only the contribution of the third generation.

It is straightforward to see that the maximal value of the rate is achieved in the case of a bottom quark running in the loop, since, due to the SM neutrino quantum
numbers, it is impossible to construct a loop with an intermediate top quark.
Taking $m_b = 4 $ GeV, the lifetime of the DM in this case can be estimated as:
\begin{equation}
\label{eq:lifetimecase1}
\tau\left(\psi \rightarrow \gamma \nu\right) \simeq 5.6 
\times 10^6\,\mbox{s}\,\, {\left(\frac{m_\psi}{7\,\mbox{keV}}\right)}^{-3}
{\left(\frac{m_{\Sigma_f}}{1\,\mbox{TeV}}\right)}^4 
{\left(\lambda \lambda^{'}\right)}^{-2}\,.
\end{equation}
By requiring a value of the lifetime of the order $10^{28}\,\mbox{s}$, as expected for the detected photon line, we obtain the condition:
\begin{equation}
\label{eq:firstmodelprediction}
\lambda \lambda^{'}\simeq 2.4 \times 10^{-11} 
{\left(\frac{m_\psi}{7\,\mbox{keV}}\right)}^{-3/2}{
\left(\frac{m_{\Sigma_f}}{1\,\mbox{TeV}}\right)}^2 
{\left(\frac{\tau\left(\psi \rightarrow \gamma \nu\right)}
{10^{28}\,\mbox{s}}\right)}^{-1/2}\,.
\end{equation}
As evident the prediction for the value of the product $\lambda \lambda^{'}$ is much higher than the one considered in~\cite{Arcadi:2013aba,Arcadi:2014tsa}. 
This is consequence of the strong sensitivity of the DM lifetime on the DM mass. 
We can determine the single values of the two couplings $\lambda$ and $\lambda^{'}$ by combining eq.~(\ref{eq:firstmodelprediction}) with the requirement of 
the correct relic DM relic density. Indeed the latter is determined by a combination of freeze-in~\cite{Hall:2009bx} and SuperWIMP~\cite{Feng:2003xh,Feng:2003uy} 
mechanisms, both relying on the decay of the scalar field into DM, as:
\begin{align}
\label{Omegah2-DM2}
& \Omega_{\rm DM}h^2=\Omega^{\rm FI}_{\rm DM}h^2+\Omega_{\rm DM}^{\rm SW}h^2\nonumber\\
& \simeq 1.09\times 10^{27}  \frac{g_{\Sigma_f}}{g_{*}^{3/2}}\frac{m_\psi}{m_{\Sigma_f}} \frac{\Gamma\left(\Sigma_f 
\rightarrow f \psi\right)}{m_{\Sigma_f}}+\frac{m_\psi}{m_{\Sigma_f}} 
Br\left(\Sigma_f \rightarrow f \psi\right)\Omega_{\Sigma_f}h^2\,, 
\end{align}
where $g_{*}$ is the number of relativistic degrees of freedom in the Early Universe at the time of DM production while $g_{\Sigma_f}$ and $\Omega_{\Sigma_f}$ 
represent, respectively, the internal degrees of freedom and the abundance at freeze-out of $\Sigma_f$. The abundance of the scalar field is fixed by interactions 
mediated by its gauge couplings and is not influenced by the couplings $\lambda$ and $\lambda^{'}$. As shown in~\cite{Arcadi:2013aba}, the contribution of the 
SuperWIMP mechanism is negligible when $m_\psi \ll m_{\Sigma_f}$. It is then possible to directly relate the value of the DM relic density to the coupling $\lambda$ as:
\begin{equation}
\label{eq:lambdafimp}
\lambda \simeq 0.8 \times 10^{-8} {\left(\frac{m_\psi}
{7\,\mbox{keV}}\right)}^{-1/2}{\left(\frac{m_{\Sigma_f}}
{1\,\mbox{TeV}}\right)}^{1/2}{\left(\frac{g_{*}}{100}\right)}^{3/4}
{\left(\frac{\Omega h^2}{0.11}\right)}^{1/2}\,.
\end{equation}
Substituting this result into eq.~(\ref{eq:firstmodelprediction}) we get:
\begin{equation}
\label{eq:lambdapfirstscenario}
\lambda^{'} \simeq 3 \times 10^{-3} {\left(\frac{m_\psi}{7\,\mbox{keV}}\right)}^{-2}
{\left(\frac{m_{\Sigma_f}}{1\,\mbox{TeV}}\right)}^{5/2} 
{\left(\frac{\tau\left(\psi \rightarrow \gamma \nu\right)}
{10^{28}\,\mbox{s}}\right)}^{-1/2}\,.
\end{equation}
We note that there is a much stronger hierarchy between the couplings $\lambda$ and $\lambda^{'}$ with 
respect to the one found in \cite{Arcadi:2013aba,Arcadi:2014tsa}. 
Indeed, in order to compensate the suppression of the decay rate due to the DM mass 
of the order of keV, we can only increase the coupling $\lambda^{'}$ 
since the coupling $\lambda$, instead, fixed by the freeze-in mechanism as in eq.~(\ref{eq:lambdafimp}), \
is a rather slowly varying function of the DM mass and is still very suppressed. 
The scalar field $\Sigma_f $ thus decays dominantly only into SM particles. The associated decay length is given by:
\begin{equation}
l_{\Sigma_f} \simeq 5.6 \times 10^{-11}\,\mbox{cm} 
{\left(\frac{m_\psi}{7\,\mbox{keV}}\right)}^{2}{\left(\frac{m_{\Sigma_f}}
{1\,\mbox{TeV}}\right)}^{-4} \left(\frac{\tau\left(\psi \rightarrow \gamma \nu\right)}
{10^{28}\,\mbox{s}}\right)\,.
\end{equation}
As a consequence the scalar field is expected to promptly decay if produced at LHC. 
We remark that, due to the dependence of eq.~(\ref{partialDecayWidth}) on the internal fermion mass, the value of $\lambda^{'}$ reported in~(\ref{eq:lambdapfirstscenario}) 
is the minimal achievable. The conclusion above hence is valid for all the realizations given in~(\ref{eq:lagrangian}).

Contrary to the scenarios discussed in~\cite{Arcadi:2013aba,Arcadi:2014tsa}, the requirement of the 
correct relic density and a Indirect Detection of DM decays for masses in the keV range 
leads to the prediction of a promptly decaying scalar field with two 
possible decay channels into a third generation quark and a neutrino or a charged lepton. Conventional constraints from LHC searches 
then apply. 
In the case of color charged scalar fields the relevant bounds come from searches of leptoquarks. The most severe limits have been at the moment set by CMS and
for the scenario under consideration masses of the scalar field below approximately 
840 GeV are excluded \cite{CMS-PAS-EXO-12-041}. 
This limit is weakened down to 740 GeV~\cite{Khachatryan:2014ura} if coupling with only third generation fermions is assumed. In the case of only 
electroweakly interacting scalar field the strongest limit come from searches of supersymmetric particles decaying into leptons and missing 
energy~\cite{Frigerio:2014ifa}. The upper limit on the mass of the scalar is then $m_{\Sigma_{\ell,e}} < 190-250$ GeV~\cite{Aad:2014vma,Khachatryan:2014qwa}.

As evident in the discussion, also for a keV DM it is possible to exploit the correlation between ID DM signal and collider searches and obtain a rather clear determination 
of the parameters of the model. On the other hand an hypothetical LHC signal would not able to discriminate our scenario from other models since its peculiar signature, 
i.e. the observation of both the decay modes of the scalar field (DM+SM and SM only) cannot be achieved. 

\section{Dark matter and Dark radiation scenario}

\begin{figure}[t]
  \centering
\begin{tikzpicture}[line width=1.5pt,rotate=90,scale=1.5]
	\draw[scalar] (-30:0.9) -- (90:0.9);
	\draw[fermion] (90:0.9) -- (210:0.9);
	\draw[fermion] (210:0.9) -- (-30:0.9);
	\begin{scope}[shift={(-115:.5)},rotate=90]
	\draw (125:.11) -- (-55:.11);
	\draw (55:.11) -- (-125:.11);
	\end{scope}
	\begin{scope}[shift={(5:.5)},rotate=30]
	\draw (125:.11) -- (-55:.11);
	\draw (55:.11) -- (-125:.11);
	\end{scope}
	\draw[fermion] (-30:0.9) -- (-30:1.8);
	\draw[fermionbar] (90:0.9) -- (90:1.8); 
	\draw[vector] (210:0.9) -- (210:1.8);
	\node at (39:.80) {$\Sigma_{q}$};
	\node at (-8:.92) {$\Sigma_{u}$};
	\node at (210:2.05) {$\gamma$};
	\node at (-30:2.00) {$\chi$};
	\node at (285:0.75) {$t_R$};
	\node at (230:0.93) {$t_L$};
	\node at (155:0.75) {$t_L$};
	\node at (90:2.00) {$\psi$};
\end{tikzpicture}
\hspace{9mm}
\begin{tikzpicture}[line width=1.5pt,rotate=90,scale=1.5]
	\draw[fermion] (-30:0.9) -- (90:0.9);
	\draw[scalar] (90:0.9) -- (210:0.9);
	\draw[scalar] (210:0.9) -- (-30:0.9);
	\begin{scope}[shift={(-115:.5)},rotate=90]
	\draw (125:.11) -- (-55:.11);
	\draw (55:.11) -- (-125:.11);
	\end{scope}
	\begin{scope}[shift={(5:.5)},rotate=30]
	\draw (125:.11) -- (-55:.11);
	\draw (55:.11) -- (-125:.11);
	\end{scope}
	\draw[fermion] (-30:0.9) -- (-30:1.8);
	\draw[fermionbar] (90:0.9) -- (90:1.8); 
	\draw[vector] (210:0.9) -- (210:1.8);
	\node at (39:.80) {$t_L$};
	\node at (-8:.92) {$t_R$};
	\node at (210:2.05) {$\gamma$};
	\node at (-30:2.00) {$\chi$};
	\node at (285:0.85) {$\Sigma_{u}$};
	\node at (235:0.95) {$\Sigma_{q}$};
	\node at (155:0.75) {$\Sigma_{q}$};
	\node at (90:2.00) {$\psi$};
\end{tikzpicture}

\caption{Diagrams contributing at one-loop to the DM 2-body decay into 
$\gamma$ and $\nu$ induced by scalar-mixing.}
\label{loopDMdecayMixScalar}
\end{figure}
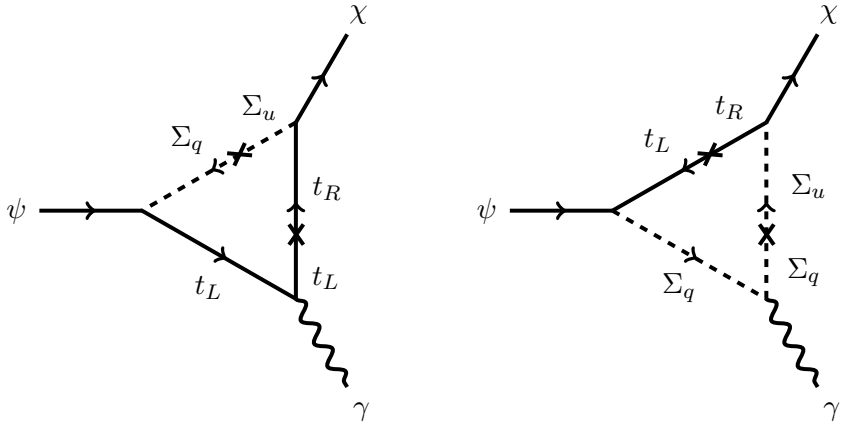

In this section we discuss an extension of the minimal decaying DM scenario in which the spectrum of BSM states is augmented with another SM singlet 
$\chi$ and the scalar spectrum is constituted by two fields, a SU(2) doublet and a singlet. The photon line is now produced by the decay 
$\psi \rightarrow \chi \gamma$ (see fig.\ref{loopDMdecayMixScalar}) and it is described by the following lagrangian:

\begin{align}
& L_{\rm eff}= \left(\lambda_{\rm L} \bar{\psi} q_L \Sigma_q^{\dagger}+\lambda_{\rm R}\bar{\psi} t_R \Sigma_u^{\dagger}\right)+h.c. \nonumber\\
& +\left(\lambda^{'}_L \bar \chi q_L \Sigma_{q}^{\dagger}+\lambda^{'}_R \bar{\chi} t_R \Sigma_{u}^{\dagger}+h.c.\right)\nonumber\\
& + \mu H \Sigma_{q} \Sigma_{u}^{\dagger}+h.c.\,.
\end{align}  

We have assigned to the two scalar fields the quantum numbers of a left handed and right-handed up-quark in order to enhance the loop function
through the top mass and possibly achieve the desired value of the DM lifetime for suppressed couplings, such that the scalar fields are long-lived at the LHC.  
Given the strong sensitivity of the DM lifetime on the SM fermion masses, we have assumed for simplicity that the two SM singlets $\chi$ and $\psi$ are coupled 
only with third generation quarks. 
The couplings of the scalar field with only SM fermions, which we have omitted for simplicity, are not relevant for the DM decay and then can be set to be of 
comparable value to the couplings governing the decay of the scalar field into DM.

In this case after electroweak symmetry breaking the fields $ \Sigma_{q}, \Sigma_{u} $ mix and
the physical fields are instead $\Sigma_{1,2}$, the eigenstates of the mass matrix:
\begin{equation}
\mathcal{M}=\left(
\begin{array}{cc}
m_{\Sigma_u}^2 & \mu v \\
\mu v & m_{\Sigma_q}^2 
\end{array}
\right),
\end{equation}
where $v$ is the v.e.v. of the Higgs field. $\mathcal{M}$ has eigenvalues
\begin{equation}
m_{\Sigma_{1,2}}^2=\frac{1}{2}\left(m_{\Sigma_q}^2+m_{\Sigma_u}^2 \mp \sqrt{\left(m_{\Sigma_q}^2-m_{\Sigma_u}^2\right)^2+4 v^2 \mu^2}\right)
\end{equation}
and is diagonalized through the generic matrix:
\begin{equation}
\eta = \left(
\begin{array}{cc}
\cos\theta & \sin\theta \\
-\sin\theta & \cos\theta
\end{array}
\right)
\end{equation}
where the mixing angle is given by
\begin{equation}
\tan 2\theta = \frac{2\mu v}{m_{\Sigma_q}^2-m_{\Sigma_u}^2}\; .
\end{equation}

The DM decay rate is then given by:
\begin{align}
& \Gamma\left(\psi \rightarrow \chi \gamma\right) = \frac{\alpha m_\psi^3}{32 \pi^4} \left(1-\frac{m_\chi^2}{m_\psi^2}\right)^3
\nonumber\\
& \left[m_t \sin\theta \cos\theta \left(\lambda_L \lambda^{'}_R-\lambda_R \lambda^{'}_L\right) 
\left(\frac{1}{m_{\Sigma_1}^2} f_1 \left(\frac{m_t^2}{m_{\Sigma_1}^2}\right)
-\frac{1}{m_{\Sigma_2}^2} f_1 \left(\frac{m_t^2}{m_{\Sigma_2}^2}\right)\right)\right. \nonumber\\
& \left. -\frac{1}{4} \left(m_\psi-m_\chi\right) 
\left(\lambda_L \lambda^{'}_L-\lambda_R \lambda^{'}_R\right)
\left(\frac{\sin^2 \theta}{m_{\Sigma_1}^2} f_2\left(\frac{m_t^2}{m_{\Sigma_1}^2}\right)
+\frac{\cos^2 \theta}{m_{\Sigma_2}^2} f_2
\left(\frac{m_t^2}{m_{\Sigma_2}^2}\right)\right)\right]^2, \nonumber\\
& \mbox{where} \;\; f_2(x)=\frac{1}{{\left(1-x\right)}^2} \left[1+x+\frac{2x}{1-x}\ln x\right]\; .
\end{align}
The term inside the square bracket in the DM decay rate is maximized for $\theta=\frac{\pi}{4}$, i.e. $m_{\Sigma_q}^2-m_{\Sigma_u}^2 = 0$, 
 and $\lambda_{L}=\lambda, \lambda_R=0, \lambda^{'}_L=0, \lambda^{'}_R=\lambda^{'}$ (or viceversa). The DM lifetime is mostly determined by the 
 contribution from the lightest eigenstate. Neglecting the mass of $\chi$ (the reason will be clarified in the following) it can be expressed as:
\begin{equation}
\tau\left(\psi \rightarrow \chi \gamma\right) \simeq 1.4 \times 10^{4}\; \mbox{s}\; 
{\left(\frac{m_\psi}{7\; \mbox{keV}}\right)}^{-3}{\left(\frac{m_{\Sigma_1}}{1\; 
\mbox{TeV}}\right)}^{4} \left(\lambda \lambda^{'}\right)^{-2}.
\end{equation}
The DM production is as well dominated by $\Sigma_1$ and $\lambda$ is again determined by 
eq.~(\ref{eq:lambdafimp}). 
The expected value of $\lambda^{'}$ from the combined requirement of reproducing the 
photon-line and the correct DM relic density is:
\begin{equation}
\label{eq:lambdachi}
\lambda^{'} \simeq 1.5 \times 10^{-4}\; {\left(\frac{m_\psi}
{7\; \mbox{keV}}\right)}^{-2}{\left(\frac{m_{\Sigma_1}}{1\; \mbox{TeV}}\right)}^{5/2}\,.
\end{equation}  
This value, although sensitively lower than the one obtained in the previous case, is still large enough to make the scalar field decay promptly at the LHC. 
Being substantially free parameters, the couplings of the scalar field with only SM fermions can be of 
comparable order as~(\ref{eq:lambdachi}) in order to allow for the 
observation of a double LHC signal. Dealing with prompt decays one has anyway to cope  already with strong limits from LHC searches. 
The most stringent ones come from searches of top squarks. Current limits allow $177 \lesssim m_{\Sigma_1} \lesssim 200$ GeV or 
$m_{\Sigma_1} > 750$ GeV~\cite{Aad:2013ija,Aad:2014bva,Aad:2014kva}. This value can be actually relaxed 
in presence of a branching ratio of 
decay into missing energy lower than 1. Indeed, in this case the scalar $\Sigma_1 $ is a mixed state and can have decays 
as both $ \Sigma_q $ and $ \Sigma_u $.


Notice also that the state $\chi$ is cosmologically stable if it is very light and might exist in sizable 
numbers contributing to the DM abundance. 
Indeed, contrary to the case of $\psi$, the value of the coupling $\lambda^{'}$ is high enough to create, 
at early stages of the history of the Universe, a thermal population of $\chi$ particles through 
decays/inverse decays of the scalar fields and $2 \rightarrow 2$ scattering with top quarks. 
$\chi$ particles then undergo a relativistic freeze-out at temperatures between 100 GeV and 1 TeV. 
In order to avoid bounds from overclosure of the Universe and structure formation we impose a very low 
mass for this new state,  $m_\chi \lesssim O(\mbox{eV})$. Such light state can affect the number of 
effective neutrinos $N_{\rm eff}$. The deviation from the SM prediction $N_{\rm eff} =3.046$ induced 
by the $\chi$ particles can be expressed as~\cite{Blennow:2012de,Valentino:2013wha}:
\begin{equation}
\Delta N_{\rm eff}= \frac{23.73}{{\left(g_{*}^s (T_d)\right)}^{4/3}}\,,
\end{equation}
where $T_d$ represents the decoupling temperature from the primordial thermal bath of the $\chi$ particles. Thanks to the rather high decoupling temperature 
we have $ g_{*}^s (T_d) \sim 100 $ due to Standard Model states and therefore 
$\Delta N_{\rm eff} \sim 0.05 $, which is compatible with the current constraints~\cite{Ade:2013zuv}.

\section{DM as sterile neutrino}

\begin{figure}[t]
  \centering
\begin{tikzpicture}[line width=1.5pt,rotate=90,scale=1.5]
	\draw[vector] (-30:0.9) -- (90:0.9);
	\draw[fermion] (90:0.9) -- (210:0.9);
	\draw[fermion] (210:0.9) -- (-30:0.9);
	\begin{scope}[shift={(90:1.2)},rotate=90]
	\draw (125:.11) -- (-55:.11);
	\draw (55:.11) -- (-125:.11);
	\end{scope}
	\draw[fermion] (-30:0.9) -- (-30:1.8);
	\draw[fermionbar] (90:0.9) -- (90:1.8); 
	\draw[vector] (210:0.9) -- (210:1.8);
	\node at (30:.76) {$W$};
	\node at (210:2.05) {$\gamma$};
	\node at (-30:2.00) {$\nu_L$};
	\node at (275:0.75) {$\ell_L$};
	\node at (100:0.99) {$\nu_L$};
	\node at (155:0.75) {$\ell_L$};
	\node at (90:2.00) {$\psi$};
\end{tikzpicture}
\hspace{9mm}
\begin{tikzpicture}[line width=1.5pt,rotate=90,scale=1.5]
	\draw[fermion] (-30:0.9) -- (90:0.9);
	\draw[vector] (90:0.9) -- (210:0.9);
	\draw[vector] (210:0.9) -- (-30:0.9);
	\begin{scope}[shift={(90:1.2)},rotate=90]
	\draw (125:.11) -- (-55:.11);
	\draw (55:.11) -- (-125:.11);
	\end{scope}
	\draw[fermion] (-30:0.9) -- (-30:1.8);
	\draw[fermionbar] (90:0.9) -- (90:1.8); 
	\draw[vector] (210:0.9) -- (210:1.8);
	\node at (30:.76) {$\ell_L$};
	\node at (210:2.05) {$\gamma$};
	\node at (-30:2.00) {$\nu_L$};
	\node at (275:0.75) {$W$};
	\node at (100:0.99) {$\nu_L$};
	\node at (155:0.75) {$W$};
	\node at (90:2.00) {$\psi$};
\end{tikzpicture} 
\caption{Diagrams contributing at one-loop to the DM 2-body decay into 
$\gamma$ and $\nu$ induced by $W$.}
\label{loopDMdecayVEV}
\end{figure}
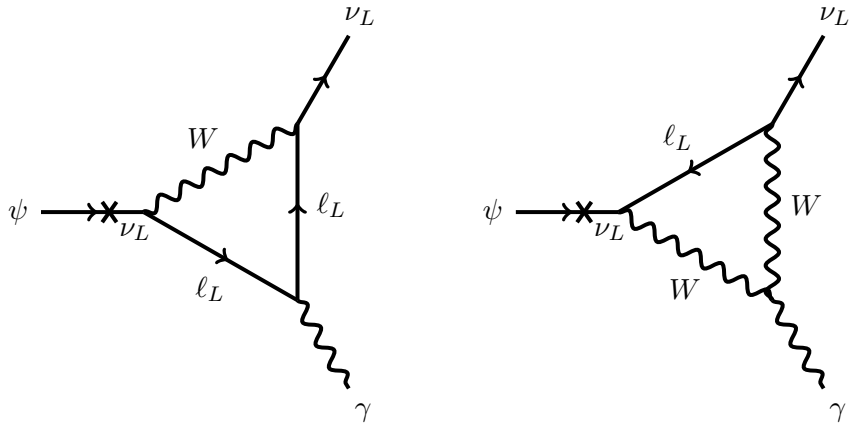

Another very simple way to reproduce the $3.55$ keV line is to allow for a coupling with the Higgs boson and a SM neutrinos of the form 
$\tilde{\lambda} \bar \psi H \ell$, where for simplicity we have suppressed generation indices, thus identifying the DM with a sterile right-handed neutrino. 
In this case the radiative decay of the DM is achieved, irrespectively of couplings and mass of the scalar field, through loops involving charged 
leptons and the $W$ boson, as shown in fig.\,\ref{loopDMdecayVEV}. 

The decay rate of the DM is given by~\cite{Boyarsky:2005us}:
\begin{equation}
\Gamma(\psi \rightarrow \nu \gamma)= 
\frac{9 \alpha G_F^2 m_\psi^5}{256\,4\pi^4}\sin^2 2\Theta\,, 
\end{equation}
where $\Theta=\frac{\tilde{\lambda}v}{m_\psi}$. 
The required value of the DM lifetime is obtained for $\sin^2 2\Theta=2-20 
\times 10^{-11}$ which corresponds to $\tilde{\lambda}\simeq10^{-13}$.

Although the scalar field is not responsible for the radiative decay of the DM, its presence is anyway 
relevant for the DM production. As already shown 
for instance in~\cite{Petraki:2007gq,Merle:2013wta,Molinaro:2014lfa,Abada:2014zra,Frigerio:2014ifa}, the decay of an extra field provides a simple and 
economical mechanism for the production of sterile neutrinos with the low value of mixing angle with active neutrinos compatible with DM Indirect Detection. 
Indeed the conventional, non-resonant, production through oscillations from active 
neutrinos, known as the Dodelson-Widrow mechanism~\cite{Dodelson:1993je}, 
can provide no more than $1 \%$ of the DM relic density for the values of the parameters accounting 
for the present $X$-ray signal, and its resonant 
enhancement~\cite{Shi:1998km} requires the presence of a very large lepton asymmetry at low temperatures, which it is not trivial to achieve 
in realistic scenarios~\cite{Shaposhnikov:2008pf}. We also remark that the distribution function of the DM arising from the decays is ``colder'' with respect 
to the two mechanisms mentioned above, thus alleviating tensions with structure formation bounds for DM candidates with mass of the order of the keV.

The LHC phenomenology depends on the size of eventual couplings between the scalar field $\Sigma_f$ with only SM fields. These kind of couplings are 
completely uncorrelated to DM phenomenology and are constrained only by the assumption that they do not contribute substantially to the DM
decay nor allow for fast proton decay. The most interesting case is therefore when their value is 
suppressed and, as could be argued for example by 
a common generation mechanisms, is comparable with the one of the other couplings either $\tilde\lambda$ 
or $\lambda$ fixed as in eq.~(\ref{eq:lambdafimp}). 
In such a case we would expect an LHC-metastable $\Sigma $ field, whose prospects of detection 
have extensively discussed in~\cite{Arcadi:2014tsa}.

\section{Conclusions}

We have here considered the possibility of reproducing the $3.55 $ keV X-ray line 
signal in few simple decaying DM scenarios. 
In the minimal realization, namely the extension of the SM with a DM Majorana fermion and a single scalar field at the TeV scale, it is possible 
to produce the right abundance of DM and obtain the correct DM lifetime for reasonable values of the couplings $\lambda $ and $\lambda'$. 
The model then predicts prompt decays of the colored $\Sigma $ scalar at LHC though the larger coupling $\lambda'$ in only Standard Model
states. Even if the second decay channel of $\Sigma $ is too suppressed to be observable at the LHC, all the parameters of the models can be 
determined by a combination of the Indirect Detection and LHC measurements and the assumption of freeze-in production for DM can
be in principle tested.

We tried to see if it is possible to lower the $ \lambda'$ coupling and enhance the possibility to measure both decay channels at the LHC.
Unfortunately this turned out to be not so simple to achieve. Enlarging the Dark Matter and $\Sigma $ 
sector to have a top particle in 
the loop, does allow for a wider range of possible $\lambda' $ couplings, but does not modify strongly the hierarchy between $ \lambda'$ and $\lambda $.
In this model DM decays into a photon and an extremely light SM singlet, which can affect the number of 
cosmological relativistic degrees 
of freedom $N_{\rm eff}$ and possibly be detected in the CMB. In this case the couplings between the $ \Sigma $ fields and purely SM fields
are not fixed by the DM lifetime and the phenomenology strongly depends on their values. 
Indeed, if they are comparable to the coupling of the scalar with the new singlet fermion $\chi $,  the scalar $ \Sigma $ could decay promptly at the 
LHC in both channels. Then it may be possible to contemporary observe prompt decays into top quark and missing energy as well as decays into 
only SM states and determine again all the parameters of the model. On the other hand, if the couplings of the $ \Sigma $ fields with only SM fields
are of the order of the freeze-in coupling $ \lambda $, the scenario will be difficult to test at colliders.

We have finally considered the case in which the DM has the additional coupling to the Higgs field and a SM neutrino, similarly to a sterile neutrino. 
The DM lifetime is determined only by the new mixing, but the coupling of DM with the SM-charged field $\Sigma $ can help in obtaining the
right DM abundance.  In this last case, if also the couplings of $ \Sigma $ to SM fields are of similar size, a  double LHC detection of $\Sigma $ 
decays through displaced vertices could be possible at LHC and disentangle this scenario from a pure sterile neutrino.

\acknowledgments

\noindent
The authors thank Asmaa Abada, Torsten Bringmann and Ninetta Saviano for their useful comments and suggestions.
G.A. thanks the Institute for Theoretical Physics of the Georg-August University G\"{o}ttingen for the warm hospitality during 
part of the completion of this work.  

\noindent
G.A. acknowledges support from the ERC advanced grant Higgs@LHC.
The authors acknowledge partial support from the European Union FP7 ITN-INVISIBLES (Marie Curie Actions, PITN-GA-2011-289442).

\bibliography{bibfile}{}
\bibliographystyle{JHEP}
\end{document}